\documentclass[reprint,superscriptaddress,nofootinbib,twocolumn,amsmath,amssymb,aps,pra]{revtex4-2}
\usepackage{amsmath,gensymb,textcomp,bm,dcolumn,eurosym,array,tabu,multirow,nicefrac,color,graphicx,upgreek}
\usepackage[colorlinks, linkcolor=blue, citecolor=blue, urlcolor=blue, breaklinks=true]{hyperref}

\usepackage{mathpazo,times} % Use a nicer font than the default

\newcommand{\bra}[1]{\langle #1 \rvert}
\newcommand{\ket}[1]{\lvert #1 \rangle}

\usepackage[subsectionbib]{bibunits}
\usepackage{graphicx}
\usepackage{epstopdf}
\usepackage{epsfig}
\usepackage{dcolumn}
\usepackage{bm,MnSymbol}
\usepackage{units}
\usepackage{psfrag}
\usepackage{color}

\newcommand{\pare}[1]{\left( #1 \right)}

\newcommand{\abs}[1]{\left\vert #1 \right\vert}
\newcommand{\cor}[1]{\left[ #1 \right]}
\newcommand{\llav}[1]{\left\lbrace #1 \right\rbrace}

\begin{document}

\title{Role of time-frequency correlations in two-photon-two-atom resonance energy transfer}

\author{Roberto de J. Le\'on-Montiel}
%\email{roberto.leon@nucleares.unam.mx}
%\thanks{\textbf{Authors to whom correspondence should be addressed:}\\ roberto.leon@nucleares.unam.mx, alfred.uren@correo.nucleares.unam.mx, and dongsh2@yahoo.com}
\affiliation{Instituto de Ciencias Nucleares, Universidad Nacional Aut\'onoma de M\'exico, Apartado Postal 70-543, 04510 Cd. Mx., M\'exico}

\author{Arturo Pedroza-Rojas}
\affiliation{Instituto de Ciencias Nucleares, Universidad Nacional Aut\'onoma de M\'exico, Apartado Postal 70-543, 04510 Cd. Mx., M\'exico}

\author{Jorge A. Peralta-\'Angeles}
\affiliation{Instituto de Ciencias Nucleares, Universidad Nacional Aut\'onoma de M\'exico, Apartado Postal 70-543, 04510 Cd. Mx., M\'exico}

%\author{\'Aulide Mart\'inez-Tapia}
%\thanks{These authors contributed equally}
%\affiliation{Instituto de Ciencias Nucleares, Universidad Nacional Aut\'onoma de M\'exico, Apartado Postal 70-543, 04510 Cd. Mx., M\'exico}

%\author{Samuel Corona-Aquino}
%\thanks{These authors contributed equally}
%\affiliation{Instituto de Ciencias Nucleares, Universidad Nacional Aut\'onoma de M\'exico, Apartado Postal 70-543, 04510 Cd. Mx., M\'exico}

%\author{Freiman Triana-Arango}
%\thanks{These authors contributed equally}
%\affiliation{Centro de Investigaciones en Optics AC, Apartado Postal 37150, Leon, Gto, Mexico}

%\author{Chenglong You}
%\affiliation{Quantum Photonics Laboratory, Department of Physics \& Astronomy, Louisiana State University, Baton Rouge, LA 70803, USA}

%\author{Rui-Bo Jin}
%\affiliation{Hubei Key Laboratory of Optical Information and Pattern Recognition, Wuhan Institute of Technology, Wuhan 430205, China}

%\author{Omar S. Maga\~na-Loaiza}
%\affiliation{Quantum Photonics Laboratory, Department of Physics \& Astronomy, Louisiana State University, Baton Rouge, LA 70803, USA}

%\author{Shi-Hai Dong}
%\email{dongsh2@yahoo.com}
%\affiliation{Research Center for Quantum Physics, Huzhou University, Huzhou 313000, China}
%\affiliation{Laboratorio de Ciencias de la Informacion Cuantica, CIC, Instituto Politecnico Nacional, UPALM, C.P 07700, CDMX, Mexico}

%\author{Alfred B. U'Ren}
%\email{alfred.uren@correo.nucleares.unam.mx}
%\affiliation{Instituto de Ciencias Nucleares, Universidad Nacional Aut\'onoma de M\'exico, Apartado Postal 70-543, 04510 Cd. Mx., M\'exico}

\begin{abstract}
Excitation energy transfer is a photophysical process upon which many chemical and biological phenomena are built. From natural small systems to synthetic multichromophoric macromolecules, energy transfer deals with the process of migration of electronic excitation energy from an excited donor to an acceptor. Although this phenomenon has been extensively studied in the past, the rapid evolution of quantum-enabled technologies has motivated the question on whether nonclassical sources of light, such as entangled photon pairs, may provide us with a better control (or enhancement) of energy transfer at the nanoscale. In this work, we provide a comprehensive study of the joint excitation of two non-interacting two-level atoms by time-frequency correlated photon pairs---whose central frequencies are not resonant with the individual particles---generated by means of spontaneous parametric down conversion (SPDC). We demonstrate that while strong frequency anti-correlation between photons guarantees a large \textcolor{black}{two-photon excitation (TPE) probability}, photons bearing a sine cardinal spectral shape exhibit a $\sim$3.8 times larger \textcolor{black}{TPE signal} than photons with a Gaussian spectrum. More importantly, we find that suppression of time-ordered excitation pathways does not substantially modify the \textcolor{black}{TPE probability} for two-photon states with a Gaussian spectral shape; whereas photons with a sine cardinal spectrum exhibit the strongest \textcolor{black}{TPE signals} when two-photon excitation pathways are not suppressed. Our results not only help elucidating the role of time-frequency correlations in resonance energy transfer with SPDC photons, but also provide valuable information regarding the optimal source to be used in its experimental implementation.         
\end{abstract}
\maketitle

\section{Introduction}
Resonance energy transfer (RET) is one of the most important and vital phenomena in nature, which triggers various fundamental molecular processes. Originally described by F\"orster \cite{forster1948}, RET deals with the relocation of energy from an initially excited donor to an acceptor. The photophysical molecular mechanisms behind RET have extensively been studied from different quantum, semiclassical and even classical perspectives \cite{allcock1998,scholes2001,jenkins2003,jang2004,Renger2009,zimanyi2010,sener2011,kuhn_book2011,roberto2013,roberto2014}. However, recent theoretical and experimental investigations describing the excitation of matter by nonclassical light sources, such as entangled pairs of photons \cite{dorfman2016,schlawin2017,oka2018-1,oka2018-2,svozilik2018-1,svozilik2018-2,burdick2018,shi2020entanglement,villabona2020,Mertenskotter:21,raymer2021entangled,landes2021quantifying,raymer2021,parzuchowski2021,tabakaev2021,landes2021,munoz2021,mikhaylov2022,cushing2022,samuel2022,tabakaev2022,cutipa2022,Chen2022-1,Chen2022-2,triana2023,aulide2023,triana2024,tobias2024}, have motivated the question on whether quantum properties of light may provide us with a better control or enhancement of energy transfer at the micro- and nano-scale \cite{schatz2019}. This has particularly been fueled by the prediction and observation of fascinating entangled two-photon absorption (ETPA) phenomena, such as two-photon-induced transparency \cite{fei1997,guzman2010}, the excitation of usually forbidden atomic transitions \cite{muthukrishnan2004}, the manipulation of quantum pathways of matter \cite{roslyak2009,raymer2013,schlawin2016,schlawin2017-1,schlawin2017-2}, and the control of molecular processes with quantum light \cite{shapiro2011,shapiro_book}. One of the most appealing features of ETPA is the linear relationship between the two-photon absorption rate and the excitation intensity \cite{javanainen1990,dayan2005,lee2006}. Such findings hold significant promise for the development of highly-efficient quantum-enabled devices where nonlinear optical phenomena can be excited at low photon fluxes \cite{schlawin2018}.  

In this work, we investigate the joint excitation of two non-interacting two-level particles (two-photon acceptor). We focus on this acceptor model system, because it is well known that efficient two-photon resonant excitation is possible with classical light sources when the two-level particles are coupled \cite{rios1980,pedrozo2012,upton_2013}. Our model considers that the excitation energy is provided by a single donor that produces time-frequency correlated (and uncorrelated) photon pairs by means of spontaneous parametric downconversion (SPDC). The central frequencies of downconverted photons are chosen to be non-resonant with the individual two-level particles. Note that in the context of two-photon excitation with classical light sources, large one-photon detuning leads to a zero \textcolor{black}{two-photon excitation (TPE)} probability for uncoupled two-level acceptors \cite{muthukrishnan2004}.

As pointed out by previous authors \cite{muthukrishnan2004,fabre2013,schatz2019}, we observe that strong frequency anti-correlation between photons guarantees a large \textcolor{black}{TPE probability}. However, by using a general SPDC excitation model, we find that photons bearing a sine cardinal joint spectral shape exhibit $\sim$3.8 times larger \textcolor{black}{TPE signals} than photons with a Gaussian spectrum. It is worth pointing out that previous investigations have shown that complex superpositions of coherent states may exhibit \textcolor{black}{TPE efficiencies} comparable to those of SPDC Gaussian photons \cite{fabre2013}. This implies that even classical approaches to joint excitation of non-interacting acceptors will show a lower efficiency than sine cardinal SPDC photons. Finally, and in stark contrast to previous works, we show that suppression of time-ordered excitation pathways does not substantially modify the \textcolor{black}{TPE probability} for two-photon states with a Gaussian joint spectrum; while photons with a sine cardinal spectrum exhibit the largest \textcolor{black}{TPE signals} when two-photon excitation pathways are not suppressed. Our results help elucidating the role of time-frequency correlations in resonance energy transfer with SPDC photons, and provides valuable information regarding the optimal source that may be used in its experimental implementation.

\section{Theoretical Framework}

\subsection{Acceptor Model}
\begin{figure}[t!]
    \centering
    \includegraphics[width = 8.5cm]{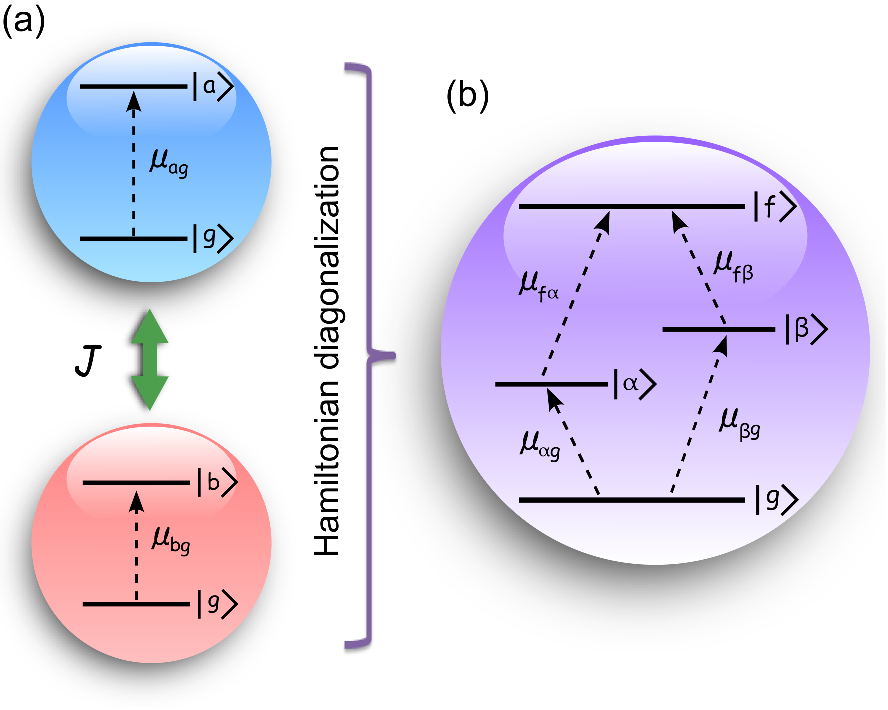}
    \caption{Schematic representation of the transformation [Eqs. (\ref{Eq3-4})-(\ref{Eq6})] that allows us to move from (a) two coupled two-level systems to a (b) single four-level system. The dashed-line, black arrows indicate all possible dipole allowed optical transitions.}
    \label{Fig1}
\end{figure}

Throughout this work, we will assume that our two-photon acceptor comprises a pair of non-interacting two-level systems, with ground and excitation frequencies $\omega_{g}$ and $\omega_{a,b}$, respectively. However, for the sake of completeness, we introduce a general two-particle model that allows us to directly apply time-dependent second-order perturbation theory to find the probability that photon pairs are absorbed by a two-particle acceptor, regardless of the coupling strength between the subsystems. Figure \ref{Fig1} shows the schematic of the transformation that allows us to move from two coupled two-level systems [Fig. 1(a)] to a single four-level system [Fig. 1(b)]. The Hamiltonian for the system shown in Fig. 1(a), represented in the $\llav{\ket{g},\ket{a},\ket{b},\ket{f}}$ basis, reads

\begin{equation} \label{Eq:2L}
H_{a} = \begin{pmatrix}
\omega_{g} & 0 & 0 & 0\\
0 & \omega_{a} & J & 0\\
0 & J & \omega_{b} & 0\\
0 & 0 & 0 & \omega_{f}\\
\end{pmatrix},
\end{equation}
where $J$ stands for the coupling strength between the two-level systems. As for the basis, note that $\ket{g}$ is the electronic state where both subsystems are in their ground states. $\ket{a}$ and $\ket{b}$ correspond to the electronic states where only one subsystem ($a$ or $b$, respectively) is excited. Finally, the state $\ket{f}$ refers to the electronic state where both transitions $a$ and $b$ are jointly excited. Note that in writing Eq. (\ref{Eq:2L}) we have assumed that the mean excitation time for each two-level system is much shorter than the lifetime of the final state. In other words, we assume that the two-level excited states have infinite lifetime.   

By diagonalizing the Hamiltonian in Eq. (\ref{Eq:2L}), we find that the coupled-particle system [Fig. 1(a)] can be represented by a single four level system [Fig. 1(b)], written in terms of the eigenstate basis $\llav{\ket{g},\ket{\alpha},\ket{\beta},\ket{f}}$, as \cite{yuen_book}
\begin{equation} \label{Eq:4L}
H_{b} = \begin{pmatrix}
\omega_{g} & 0 & 0 & 0\\
0 & \omega_{\alpha} & 0 & 0\\
0 & 0 & \omega_{\beta} & 0\\
0 & 0 & 0 & \omega_{f}\\
\end{pmatrix},
\end{equation}
where 
\begin{equation}
\ket{\alpha} = -\text{sin}\theta\ket{a} +  \text{cos}\theta\ket{b}; \;\; \ket{\beta} = \text{cos}\theta\ket{a} + \text{sin}\theta\ket{b}. \label{Eq3-4}
\end{equation}
The new (eigen)frequencies are given by $\omega_{\alpha} = \bar{\omega} - \delta\text{sec}\pare{2\theta}$, and $\omega_{\beta} = \bar{\omega} + \delta\text{sec}\pare{2\theta}$, with $\bar{\omega}=\pare{\omega_{a}+\omega_{b}}/2$, $\text{tan}\pare{2\theta}=J/\delta$, and $\delta = \pare{\omega_{a}-\omega_{b}}/2$. Note that in both representations, the conservation of energy is satisfied by the condition: $\omega_{f}=\omega_{a}+\omega_{b} = \omega_{\alpha}+\omega_{\beta}$. To complete the acceptor model, we can transform the transition dipoles of individual two-level systems by means of the following expressions:
\begin{eqnarray}
\begin{pmatrix}
\mu_{\alpha g} \\
\mu_{\beta g}\\
\end{pmatrix} &=& \begin{pmatrix}
-\sin\theta & \cos\theta \\
\cos\theta & \sin\theta\\
\end{pmatrix} \begin{pmatrix}
\mu_{a g} \\
\mu_{b g}\\
\end{pmatrix}, \label{Eq5}\\
\begin{pmatrix}
\mu_{f \alpha} \\
\mu_{f \beta}\\
\end{pmatrix} &=& \begin{pmatrix}
\cos\theta & -\sin\theta \\
\sin\theta & \cos\theta\\
\end{pmatrix} \begin{pmatrix}
\mu_{a g} \\
\mu_{b g}\\
\end{pmatrix}. \label{Eq6}
\end{eqnarray}
Note that by using Eqs. (\ref{Eq3-4})-(\ref{Eq6}), we can easily move from a picture where the acceptor is described by a coupled-particle, four-level system to two uncoupled two-level systems, simply by changing the coupling coefficient, $J$, from a finite non-zero value to zero. Thus, for the sake of simplicity, in what follows we will use the four-level configuration to describe our acceptor.

\subsection{Donor Model}
Having defined our acceptor model, we now assume that two-photon RET involves the absorption of correlated photon pairs produced by a single donor. Previous authors have mostly focused on donors where time-frequency correlated photons are produced in cascade emission from a three level system \cite{muthukrishnan2004,fabre2013,schatz2019}. However, with the goal of investigating whether time-ordered interaction pathways (which are intrinsic to the cascade emission) and a specific two-photon spectral shape are needed for efficient RET, we assume that correlated photon pairs are produced by spontaneous parametric down-conversion (SPDC).

In general, the state describing the two-photon emission of a SPDC donor can be written as $\left|\Psi \right\rangle  = \int d\Omega_{s}d\Omega_{i}\Phi\left(\Omega_{s},\Omega_{i}\right)\left|\omega_{s}^{0}+\Omega_{s}\right\rangle _{s}\left|\omega_{i}^{0}+\Omega_{i}\right\rangle _{i}$, where $s$ and $i$ stand for the signal and idler photonic modes, $\Omega_j = \omega_{j}-\omega_{j}^{0}\;\;(j=s,i)$ are the frequency deviations from the central frequencies $\omega_{j}^{0}$, and $\Phi\pare{\Omega_s,\Omega_i}$ is the joint spectral amplitude, or mode function, which describes the type of correlations and spectral shape of the two-photon state.  

Without loss of generality, we assume that photon pairs are produced by means of type-II SPDC. In this process, photons with orthogonal polarizations are generated in a rectangularly-shaped, second-order nonlinear crystal of length $L$, when pumped by a Gaussian pulse with temporal duration $T_{p}$. The normalized mode function for such configuration reads \cite{torres}
\begin{eqnarray}
\Phi_{\text{SD}}\left(\Omega_{s},\Omega_{i}\right) &=& \sqrt{\frac{4T_{e}T_{p}}{\pi\sqrt{2\pi}}}E_{p}\left(\Omega_{s}+\Omega_{i}\right)\text{sinc}\left[T_{e}\left(\Omega_{i}-\Omega_{s}\right)\right] \nonumber \\ 
&\times& e^{i\frac{LN_{p}}{2}\left(\Omega_{s}+\Omega_{i}\right)}e^{-i\frac{LN_{s}}{2}\Omega_{i}}e^{-i\frac{LN_{i}}{2}\Omega_{s}}, \label{Eq:Phi_SD}
\end{eqnarray}
with $E_{p}\left(\Omega_{s}+\Omega_{i}\right)=\exp\cor{-T_{p}^{2}\left(\Omega_{s}+\Omega_{i}\right)^{2}}$. The correlation (entanglement) time between photons is given by $T_{e} = \pare{N_{s}-N_{i}}L/4$, and $N_{j}$ $\pare{j=i,s,p}$ are the inverse group velocities of the idler, signal and pump fields, respectively. In writing Eq. (\ref{Eq:Phi_SD}), we have made use of the group velocity matching condition $N_{p}=\pare{N_{i}+N_{s}}/2$ \cite{keller1997,hendrych2007}, which eases the tuning of the two-photon frequency correlations. Note that the last two terms in Eq. (\ref{Eq:Phi_SD}) indicate that the signal and idler photons exit the nonlinear crystal at different times. This so-called photon walk-off creates a time-ordering in the interaction with the acceptor. In other words, there is only one excitation pathway in which the signal photon interacts first with the acceptor. Because of this, we will refer Eq. (\ref{Eq:Phi_SD}) as to \emph{sine cardinal distinguishable photon pairs}. We can recover the missing pathway---namely the idler photon interacting with the acceptor first---by interchanging the polarization of the photons (once they leave the crystal) and making them traverse a similar crystal of length $L/2$. In this scenario, the state of the \emph{sine cardinal indistinguishable photon pairs} takes the form \cite{roberto_spectral_shape}
\begin{eqnarray}
\Phi_{\text{SI}}\left(\Omega_{s},\Omega_{i}\right) &=& \sqrt{\frac{4T_{e}T_{p}}{\pi\sqrt{2\pi}}}E_{p}\left(\Omega_{s}+\Omega_{i}\right) \nonumber \\ 
&\times& \text{sinc}\left[T_{e}\left(\Omega_{i}-\Omega_{s}\right)\right] e^{i\frac{LN_{p}}{2}\left(\Omega_{s}+\Omega_{i}\right)}. \label{Eq:Phi_SI}
\end{eqnarray}

We can further explore other types of two-photon mode functions. Of particular interest are those where the spectrum of the photons is restricted by using a Gaussian spectral filter \cite{roberto_spectral_shape}. In this situation, both the \emph{distinguishable and indistinguishable Gaussian photon pairs} are described, respectively, by
\begin{eqnarray}
\Phi_{\text{GD}}\left(\Omega_{s},\Omega_{i}\right) &=& \sqrt{\frac{4\alpha T_{e} T_{p}}{\pi}}E_{p}\left(\Omega_{s}+\Omega_{i}\right)e^{-\alpha^2T_{e}^2\left(\Omega_{i}-\Omega_{s}\right)^2} \nonumber \\ 
&\times& e^{i\frac{LN_{p}}{2}\left(\Omega_{s}+\Omega_{i}\right)}e^{-i\frac{LN_{s}}{2}\Omega_{i}}e^{-i\frac{LN_{i}}{2}\Omega_{s}}, \label{Eq:Phi_GD} \\
&\hspace{-43mm}\text{and}& \nonumber \\
\Phi_{\text{GI}}\left(\Omega_{s},\Omega_{i}\right) &=& \sqrt{\frac{4\alpha T_{e} T_{p}}{\pi}} E_{p}\left(\Omega_{s}+\Omega_{i}\right) \nonumber \\ 
&\times& e^{-\alpha^2T_{e}^2\left(\Omega_{i}-\Omega_{s}\right)^2} e^{i\frac{LN_{p}}{2}\left(\Omega_{s}+\Omega_{i}\right)}, \label{Eq:Phi_GI}
\end{eqnarray}
where, for the sake of comparison, we have made used of the approximation \cite{clara2008}: $\text{sinc}\pare{\tau\omega}\approx \exp\cor{-\alpha^2\tau^2\omega^2}$, with $\alpha=0.455$. Equations (\ref{Eq:Phi_GD}) and (\ref{Eq:Phi_GI}) are particularly relevant for this work, as they are able to describe fully uncorrelated photon pairs in the case where $T_{e}=T_{p}/ \alpha$. It is worth pointing out that the condition $T_{e}<T_{p}/ \alpha$ indicates anti-correlation between photons; whereas correlated photon pairs are produced when $T_{e}>T_{p}/ \alpha$. These conditions apply to both sine cardinal and Gaussian photon pairs, with the difference that at $T_{e}=T_{p}/ \alpha$, sine cardinal photons are referred to as quasi-uncorrelated photon pairs \cite{roberto_spectral_shape}. In what follows, we will use these results to elucidate the role of frequency correlations in enhancing two-photon resonance energy transfer.

\vspace{5mm}

\subsection{Donor-Acceptor Interaction Model}
We now describe the interaction between our SPDC donor and the two-particle acceptor. We assume that the interaction is dipole-mediated and thus it is described by the time-dependent electric dipole Hamiltonian $\hat{H}\pare{t} = \hat{\mu}\pare{t}\hat{E}^{\pare{+}}\pare{t}$, where $\hat{\mu}\pare{t}$ is the dipole-moment operator and $\hat{E}^{\pare{+}}\pare{t}$ is the positive-frequency part of the electric-field operator, which for the particular case of two-photon interaction can be written as $\hat{E}^{\pare{+}}\pare{t} = \hat{E}_{1}^{\pare{+}}\pare{t} + \hat{E}_{2}^{\pare{+}}\pare{t}$, with $\hat{E}_{1,2}^{\pare{+}}\pare{t} = \int d\omega_{1,2}\sqrt{\frac{\hbar\omega_{1,2}}{4\pi\epsilon_{0}cA}}\hat{a}\pare{\omega_{1,2}}e^{-i\omega_{1,2}t}$. Here, $c$ is the speed of light, $\epsilon_{0}$ is the vacuum permittivity, $A$ is the effective area of the field and $\hat{a}\pare{\omega_{1,2}}$ is the annihilation operator of a photonic frequency mode with frequency $\omega_{1,2}$. Note that, for the sake of simplicity, the specific spatial shape (assumed as constant) and polarization (assumed to be aligned with the acceptor dipole moments) are not explicitly written.    

We consider that initially, the acceptor is in its ground state $\ket{g}$. The probability that the two-particle acceptor experiences a \textcolor{black}{TPE event}, thus driving it to its final state $\ket{f}$, is given by second-order time-dependent perturbation theory as \cite{perina1998,montiel2019} 
\begin{equation}\label{Eq:prob}
\textcolor{black}{P_{\text{TPE}}} = \abs{\frac{1}{\hbar^2}\int_{-\infty}^{\infty}\int_{-\infty}^{t_2}dt_2dt_1 \mathcal{D}(t_1, t_2) \mathcal{E}(t_1, t_2)}^{2},
\end{equation}
where
\begin{eqnarray}
    \mathcal{D}(t_1, t_2) &=& \sum_{j=\alpha,\beta}\mu_{fj}\mu_{jg}e^{-i\pare{\omega_j-\omega_f}t_2}e^{-i\pare{\omega_g-\omega_j}t_1}, \label{Eq:dipole} \\
    \mathcal{E}(t_1, t_2) &=& \bra{0}\hat{E}_{2}^{\pare{+}}\pare{t_{2}}\hat{E}_{1}^{\pare{+}}\pare{t_1}\ket{\Psi}  \nonumber \\
    & & \hspace{5mm}+ \bra{0}\hat{E}_{1}^{\pare{+}}\pare{t_{2}}\hat{E}_{2}^{\pare{+}}\pare{t_1}\ket{\Psi},  \label{Eq:field}
\end{eqnarray}
with $\mu_{fj}=\bra{f}\hat{\mu}\ket{j}$ and $\mu_{jg}=\bra{j}\hat{\mu}\ket{g}$ being the transition matrix elements of the dipole-moment operator. Note that in Eq. (\ref{Eq:field}) we have kept only the terms in which one photon from each mode (signal and idler) contributes to the overall two-photon excitation of the acceptor. 

\section{Results and discussion}
At this point, we are ready to test the efficiency of energy transfer from the SPDC donor to the two-particle acceptor for the four two-photon states described above. We thus start by substituting Eqs. (\ref{Eq:Phi_SD})-(\ref{Eq:Phi_GI}) into Eq. (\ref{Eq:prob}) to find that the \textcolor{black}{TPE probability} for each of the mode functions reads
\begin{widetext}
\begin{eqnarray}
\textcolor{black}{P_{\text{TPE}}^{(\text{SD})}} & = &  \frac{T_{p}}{T_{e}}\frac{\sqrt{2\pi}\omega_{i}^{0}\omega_{s}^{0}}{2\hbar^{2}\epsilon_{0}^{2}c^{2}A^{2}}\exp\left[-2T_{p}^{2}\left(\omega_{s}^{0}+\omega_{i}^{0}-\omega_{f}\right)^{2}\right]\left|\sum_{j=\alpha,\beta}\mu_{fj}\mu_{jg}\frac{\sin\left[\left(\omega_{j}-\omega_{i}^{0}\right)2T_{e}\right]}{\omega_{j}-\omega_{i}^{0}} e^{-i\pare{\omega_{j}-\omega_{i}^{0}}2T_{e}} \right|^{2}, \label{Eq:P_SD}\\
\textcolor{black}{P_{\text{TPE}}^{(\text{SI})}} &=& \frac{T_{p}}{T_{e}}\frac{\sqrt{2\pi}\omega_{i}^{0}\omega_{s}^{0}}{8\hbar^{2}\epsilon_{0}^{2}c^{2}A^{2}}\exp\left[-2T_{p}^{2}\left(\omega_{s}^{0}+\omega_{i}^{0}-\omega_{f}\right)^{2}\right]\left|\sum_{j=\alpha,\beta}\mu_{fj}\mu_{jg}\left[\frac{1-e^{-i\left(\omega_{j}-\omega_{i}^{0}\right)2T_{e}}}{\omega_{j}-\omega_{i}^{0}}+\frac{1-e^{-i\left(\omega_{j}-\omega_{s}^{0}\right)2T_{e}}}{\omega_{j}-\omega_{s}^{0}}\right]\right|^{2}, \label{Eq:P_SI} \\
\textcolor{black}{P_{\text{TPE}}^{(\text{GD})}} &=& \frac{\alpha T_{p}T_{e}\pi\omega_{i}^{0}\omega_{s}^{0}}{4\hbar^{2}\epsilon_{0}^{2}c^{2}A^{2}}\exp\left[-2T_{p}^{2}\left(\omega_{s}^{0}+\omega_{i}^{0}-\omega_{f}\right)^{2}\right]\left|\sum_{j=\alpha,\beta}\mu_{fj}\mu_{jg}\left\{ \mathbb{F}_{-}\left[\pare{\omega_{j}-\omega_{i}^{0}}\alpha 2T_{e}\right]+\mathbb{F}_{+}\left[\pare{\omega_{j}-\omega_{s}^{0}}\alpha 2T_{e}\right]\right\} \right|^{2}, \label{Eq:P_GD}\\
\textcolor{black}{P_{\text{TPE}}^{(\text{GI})}} & = & \frac{\alpha T_{p} T_{e}\pi\omega_{i}^{0}\omega_{s}^{0}}{4\hbar^{2}\epsilon_{0}^{2}c^{2}A^{2}}\exp\left[-2T_{p}^{2}\left(\omega_{s}^{0}+\omega_{i}^{0}-\omega_{f}\right)^{2}\right]\left|\sum_{j=\alpha,\beta}\mu_{fj}\mu_{jg}\left\{ F\left[\pare{\omega_{j}-\omega_{i}^{0}}\alpha 2T_{e}\right]+F\left[\pare{\omega_{j}-\omega_{s}^{0}}\alpha 2T_{e}\right]\right\} \right|^{2}, \label{Eq:P_GI}
\end{eqnarray}
\end{widetext}
where $\mathbb{F}_{\mp}\pare{\xi}=e^{\mp i\xi}\left\{ F\left(\xi\right)+i2e^{-\xi^{2}}\left[\text{Erfi}\left(\xi\right)-\text{Erfi}\left(\xi\pm\frac{i}{2}\right)\right]\right\}.
$
In all cases, we have displaced the ground state energy to zero, i.e., $\omega_{g}=0$. Erfi$(\xi)$ refers to the imaginary error function and $F\pare{\xi}$ is the so-called plasma dispersion function, which is defined as \cite{nakanishi,plasma_book}: $ F\left(\xi\right)=e^{-\xi^{2}}\left(1+\frac{2i}{\sqrt{\pi}}\int_{0}^{\xi}e^{y^{2}}dy\right)$.

\begin{figure}[b!]
    \centering
    \includegraphics[width = 8.4 cm]{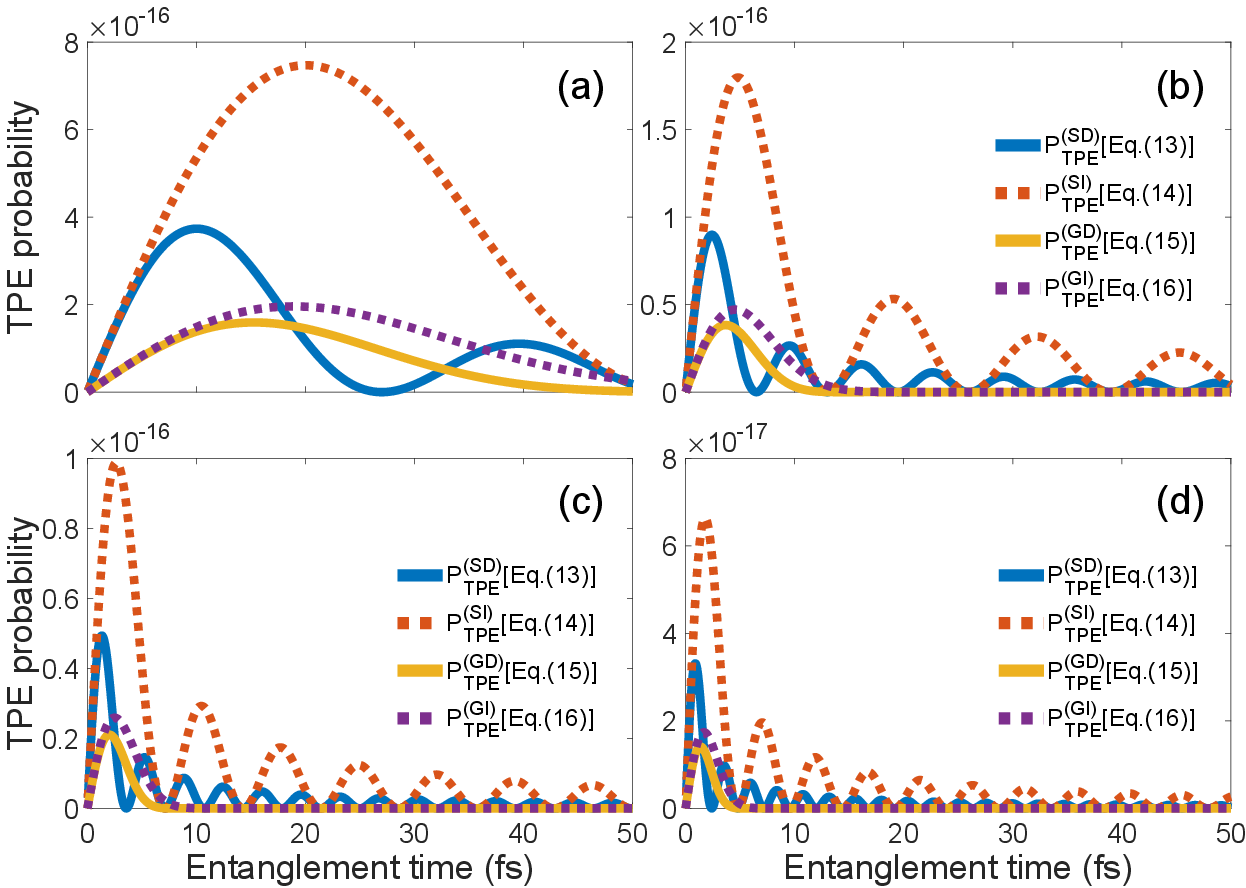}
    \caption{\textcolor{black}{Two-photon excitation (TPE) probability for distinguishable (solid lines) and indistinguishable (dotted lines) sine cardinal and Gaussian photon pairs. The parameters for our simulations are as follows: The coupling strength between acceptors is set to $J=0$. The excitation wavelengths of individual two-level systems are chosen to be (a) $\lambda_{a}=800$ nm and $\lambda_{b}=820$ nm, (b) $\lambda_{a}=770$ nm and $\lambda_{b}=854$ nm, (c) $\lambda_{a}=740$ nm and $\lambda_{b}=894$ nm, and (d) $\lambda_{a}=710$ nm and $\lambda_{b}=942$ nm. In all cases the transition dipole moments are set to $\mu_{ag}=\mu_{bg}=1$ debye. The SPDC photons are assumed to be produced at a degenerate wavelength of $810$ nm, with a pump time-duration $T_{p}=350$ fs. The effective area of the photon field is taken to be $A=1\;\mu\text{m}^{2}$.}}
    \label{Fig2}
\end{figure}

\textcolor{black}{Figure \ref{Fig2} shows the \textcolor{black}{TPE probability} for distinguishable (solid lines) and indistinguishable (dotted lines) sine cardinal and Gaussian photons. Our simulations are performed assuming that uncoupled acceptors ($J=0$) possess excitation wavelengths of (a) $\lambda_{a}=800$ nm and $\lambda_{b}=820$ nm, (b) $\lambda_{a}=770$ nm and $\lambda_{b}=854$ nm, (c) $\lambda_{a}=740$ nm and $\lambda_{b}=894$ nm, and (d) $\lambda_{a}=710$ nm and $\lambda_{b}=942$ nm. These values are chosen with the aim of investigating the behavior of the TPE signal when the excitation wavelengths of the particles are near [Fig. \ref{Fig2}(a)] and very far [Fig. \ref{Fig2}(d)] from resonance with the SPDC photons. Note that in all cases their excitation energies satisfy the two-photon resonance condition, namely $\omega_{a} + \omega_{b} = \omega_{s}^{0}+\omega_{i}^{0}$. The transition dipole moments for individual two-level systems are assumed, without loss of generality, to be $\mu_{ag}=\mu_{bg}=1$ debye. As for the SPDC donor, we consider that it emits degenerate photons (equivalent results are found with non-degenerate photon pairs) at a wavelength of $\lambda_{0}=\lambda_{s}^{0}=\lambda_{i}^{0}=810$ nm. Motivated by recent experimental photon-pair TPE experiments \cite{varnavski2023}, we set the pump duration to $T_{P}=350$ fs. Finally, the effective area of the photon field is taken to be $A=1\;\mu\text{m}^{2}$.} 

\textcolor{black}{One important difference to note between the TPE signals shown in Fig. \ref{Fig2} is the nonmonotic behavior of sine cardinal photons. This implies that even though anti-correlation (broad photon spectrum) enhances the energy transfer from the donor to the uncoupled acceptors, care needs to be taken when setting the entanglement time of a sine cardinal SPDC source. In addition, we find that suppression of time-ordered excitation pathways does not considerably modify the TPE signals for Gaussian photons -- in all cases, the maximum signal for indistinguishable Gaussian photons is $\sim 1.23$ times larger than distinguishable Gaussian photons. In striking contrast, we observe that the use of indistinguishable sine cardinal photons results in a $\sim3.81$ times larger TPE probability value with respect to the maximum value obtained for Gaussian photons. Interestingly, we find that suppression of one of the excitation pathways of sine cardinal photons leads to one half of the indistinguishable-photon TPE signal. Finally, when comparing different two-photon states, we find that the maximum TPE signal for anti-correlated sine cardinal photons is $\sim70$, $\sim550$, $\sim445$, and $\sim391$ times larger than the quasi-uncorrelated case for Figures 2(a)-(d), respectively. It is worth remarking that this value is obtained numerically by setting a range for the entanglement time $T_{e}\in\llav{0,800}$ fs, with a time-resolution $\Delta T_{e}=0.8$ fs. In the case of uncorrelated Gaussian photons, the TPE signal rapidly decreases below our numerical error, thus making it negligible when compared to the maximum value possible. Contrary to common belief, our results show that two-photon resonance energy transfer does not only depend on strong anti-correlations between photons but also on the spectral shape of such correlations. More importantly, in the particular case of sine cardinal photons, the most efficient two-photon RET can only be observed when time-ordered excitation pathways are preserved.}

\begin{figure}[t!]
    \centering
    \includegraphics[width = 8.7 cm]{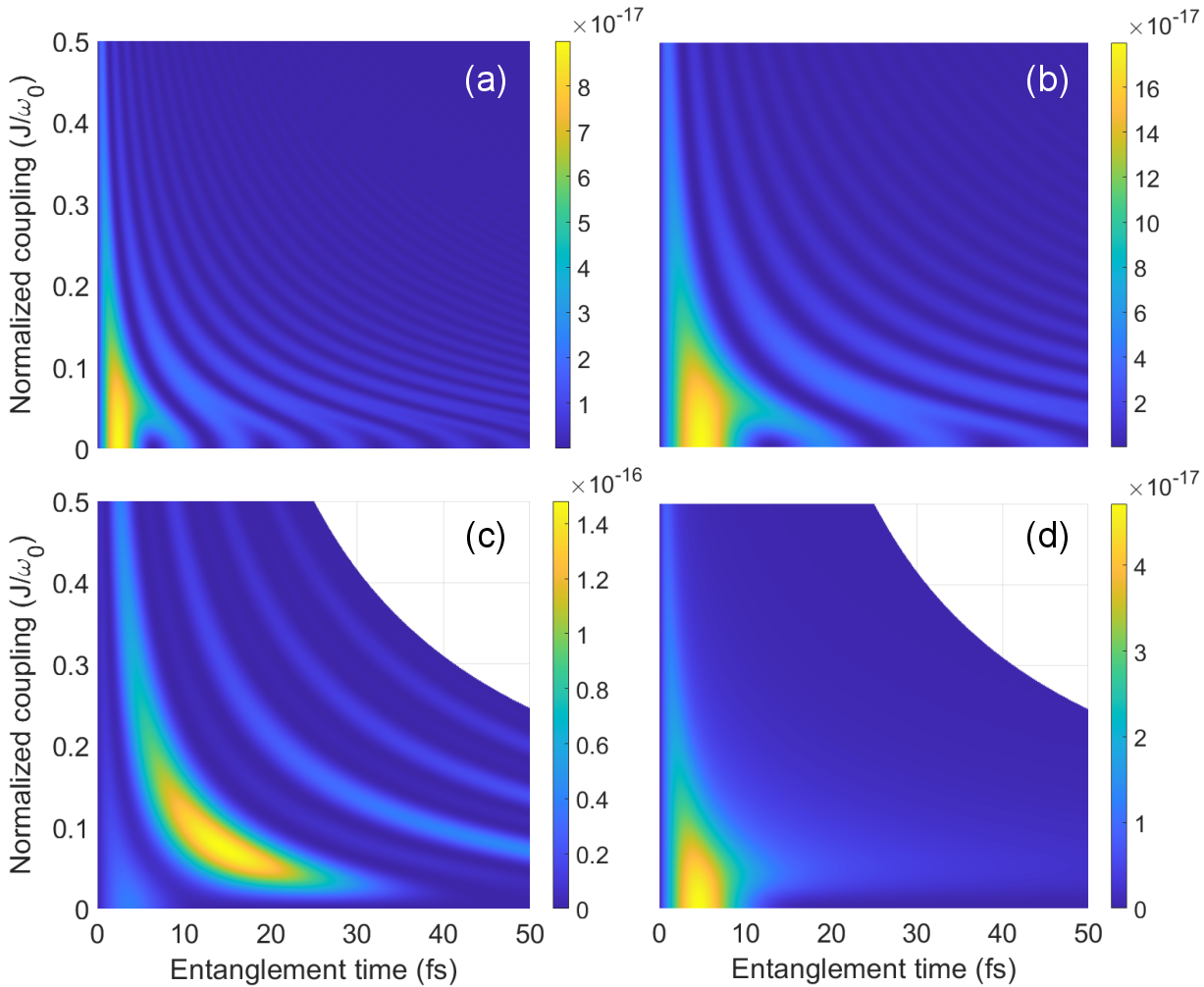}
    \caption{\textcolor{black}{Two-photon excitation (TPE) probability as a function of the entanglement time ($T_{e}$) and the coupling strength (J), normalized to the degenerate frequency of the SPDC photons, namely $\omega_{0}=2\pi c/(810\;\text{nm})$ for (a) sine-cardinal distinguishable [Eq. (13)] and (b) sine-cardinal indistinguishable [Eq. (14)] photons, as well as (c) Gaussian distinguishable [Eq. (15)] and (d) Gaussian indistinguishable [Eq. (16)] SPDC photons. The excitation wavelengths of individual two-level systems are chosen to be $\lambda_{a}=770$ nm and $\lambda_{b}=854$ nm. The individual transition dipole moments are set to $\mu_{ag}=\mu_{bg}=1$ debye. The time duration for the pump is taken to be $T_{p}=350$ fs and the effective area of the photon field is $A=1\;\mu\text{m}^{2}$. Note that the cut regions in the Gaussian photon TPE signals (bottom row) correspond to numerical-inaccessible parameter regions, where the products of exponential and Error functions create undefined operations.}}
    \label{Fig3}
\end{figure}

\textcolor{black}{For the sake of completeness, we show the behavior of the TPE signals [Eqs. (\ref{Eq:P_SD})-(\ref{Eq:P_GI})] when the coupling between acceptors is non-zero. Figure \ref{Fig3} shows the TPE signals as a function of the entanglement time ($T_{e}$) and the coupling strength (J), normalized to the degenerate frequency of the SPDC photons, namely $\omega_{0}=2\pi c/(810\;\text{nm})$. In our simulations, we consider the case where the excitation wavelengths of individual particles are given by $\lambda_{a}=770$ nm and $\lambda_{b}=854$ nm. Other combinations offer equivalent results. We can see from Fig. \ref{Fig3} that when the two-level systems are coupled, the frequency splitting introduced by the coupling strength (see the expressions for $\omega_{\alpha}$ and $\omega_{\beta}$) creates the necessity for spectrally broader photons (as the coupling increases) to successfully excite the two-particle acceptor. Interestingly, we observe that distinguishable Gaussian-shaped frequency-anti-correlated photons exhibit the strongest TPE signal of all [see, Fig. \ref{Fig3}(c)]. This implies that in the case where the particles are coupled, and interacting with SPDC photons, distinguishable Gaussian-shaped photons may offer the best avenue for increasing the efficiency of two-photon energy transfer.}

\section{Conclusion}
In summary, we have provided a thorough analysis of resonance energy transfer from a SPDC donor to an acceptor comprising two non-interacting two-level systems---whose excitation frequencies are not resonant with the central frequencies of the down-converted photons. We have found that, as pointed out by previous authors, frequency anti-correlation between photons guarantees a large (as compared to uncorrelated photons) two-photon energy transfer. However, the shape of spectral correlations are also relevant for the \textcolor{black}{TPE process} to take place efficiently. In particular, we have found that photons bearing a sine cardinal joint spectral shape may exhibit a $\sim 3.8$ times larger \textcolor{black}{TPE signal} than photons with a Gaussian spectrum. More importantly, we found that while suppression of time-ordered excitation pathways does not substantially modify the \textcolor{black}{TPE probability} for two-photon states with a Gaussian joint spectrum, photons with a sine cardinal spectrum exhibit the strongest \textcolor{black}{TPE signal} of all, when two-photon excitation pathways are not suppressed. \textcolor{black}{We would like to remark that although anti-correlated sine cardinal photons could be obtained by means of coherent control \cite{roslyak2009, fabre2013}, such process requires the use of pulsed laser sources together with fine-tuned femtosecond pulse-shaping equipment. The attractive feature of SPDC photons is that one can obtain such frequency correlations directly by pumping a nonlinear crystal (whose properties can be tuned by temperature or electric fields) with a continuous-wave laser source. This allows one to replicate pulsed laser experiments without phase modulation using low-power, compact devices \cite{Harper2023}}. The results presented here not only help elucidating the role of time-frequency correlations and photon time-(in)distinguishability in two-photon resonance energy transfer, but also shows that SPDC light offers multiple control knobs through which energy transfer can be manipulated at the nanoscale. 

\section*{Acknowledgments}
This work was supported by DGAPA-UNAM under the project UNAM-PAPIIT IN101623. J.A.P.-A. thanks financial support through a postdoctoral fellowship from CONAHCYT. We thank Juan P. Torres (ICFO) and Ralph Jimenez (JILA) for useful discussions. 

\vspace{3mm}

\bibliography{Bib_ETPA}

\end{document}